\documentclass[twoside]{IEEEtran}
\PassOptionsToPackage{ruled, linesnumbered}{algorithm2e}
\usepackage[latin9]{inputenc}
\usepackage{color}
\usepackage{algorithm2e}
\usepackage{amsmath}
\usepackage{amsthm}
\usepackage{amssymb}
\usepackage{graphicx}
\usepackage{cite}
\allowdisplaybreaks 
\makeatletter
\theoremstyle{plain}
\newtheorem{thm}{\protect\theoremname}

\usepackage{balance}
\usepackage{pgfplots}
\usepackage{pgfplotstable}
\usepackage{tikz}
\usetikzlibrary{calc}
\usetikzlibrary{arrows.meta}
\usetikzlibrary{patterns}

\usepackage{comment}

\newcommand{\trans}{^{\mathsf{T}}}

\makeatother

\providecommand{\theoremname}{Theorem}

\begin{document}
\title{\LARGE{A Max-Min Task Offloading Algorithm for
Mobile Edge Computing Using Non-Orthogonal Multiple Access}}
\author{Vaibhav Kumar,~\IEEEmembership{Member,~IEEE}, Muhammad Fainan Hanif,\\
Markku Juntti, \IEEEmembership{Fellow,~IEEE}, and Le-Nam Tran,~\IEEEmembership{Senior Member,~IEEE}\thanks{Vaibhav Kumar and Le-Nam Tran are with the School of Electrical and Electronic
Engineering, University College Dublin, D04 V1W8 Dublin, Ireland.
(email: vaibhav.kumar@ieee.org, nam.tran@ucd.ie).}
\thanks{Muhammad Fainan Hanif is with the Institute of Electrical, Electronics and Computer
Engineering, University of the Punjab, Lahore, Pakistan. (email: mfh21@uclive.ac.nz).}	
\thanks{Markku Juntti is with the Centre for Wireless Communications, University
of Oulu, 90014 Oulu, Finland (e-mail: markku.juntti@oulu.fi).}}
\maketitle
\begin{abstract}
To mitigate the computational power gap between the network core and edges, mobile edge computing (MEC) is poised to play a fundamental role in future generations of wireless networks. In this correspondence, we consider a non-orthogonal multiple access (NOMA) transmission model to maximize the worst task to be offloaded among all users to the network edge server. A provably convergent and efficient algorithm is developed to solve the considered non-convex optimization problem for maximizing the minimum number of offloaded bits in a multi-user NOMA-MEC system. Compared to the approach of optimized orthogonal multiple access (OMA), for given MEC delay, power and energy limits, the NOMA-based system considerably outperforms its OMA-based counterpart in MEC settings. Numerical results demonstrate that the proposed algorithm for NOMA-based MEC is particularly useful for delay sensitive applications. 
\end{abstract}


\section{Introduction}
\IEEEPARstart{N}{on}-orthogonal multiple access (NOMA) has gained tremendous popularity in the past few years as a potential scheme to facilitate massive connectivity, as well as to support low-latency communications. On the other hand, mobile edge computing (MEC) is being considered as a key enabler for the communication among devices with low data-processing capability, where such devices offload their computationally-intensive tasks to an edge server. The superiority of NOMA-based MEC has been proven over its orthogonal multiple access (OMA) based counterparts, in terms of energy and/or delay minimization.

In this context, the problem of energy consumption minimization in NOMA-MEC system was explored in~\cite{Ansari,Ding_MEC,FangFang,Daniel_OMA-NOMA}. On the other hand, the authors in~\cite{Ding-MEC1,Fang-MEC,Dobre,Sherman,DelayMinimization_Sherman} considered the problem of delay/latency minimization in NOMA-MEC system. More interestingly, in~\cite{StackelbergGame}, the optimization problem for a two-user NOMA-MEC system was formulated as a Stackelberg game where the users tend to minimize their energy consumption, whereas the MEC server attempts to minimize the task execution time. The problem of system cost minimization in an ultra-dense NOMA-MEC network was discussed in~\cite{UltraDense}. The problem of minimizing a weighted sum of energy consumption and latency in a NOMA-MEC system was considered in~\cite{Rev2-1} and an efficient solution was obtained using deep reinforcement learning (DRL) approach. The problem of minimizing the total energy consumption in a NOMA-MEC system using DRL in the context of industrial internet of things was presented in~\cite{Rev2-2}.

In the case of a NOMA-MEC system, the users often partition their computation tasks in two categories: low-complexity and high-complexity tasks. The low-complexity tasks are executed locally, while the high-complexity tasks are offloaded to the MEC server. Therefore, different from the existing literature, we consider the problem of maximizing the minimum number of bits (of the high-complexity tasks) offloaded by users to an MEC server using NOMA by constraining the total energy, transmission power and the offloading delay within given limits. In this context,~Huang~\emph{et al.} studied a max-min computation efficiency problem in NOMA-MEC system~\cite{Rev1-1}, and a similar problem in a millimeter wave settings was explored in~\cite{Rev1-2}. In order to improve the physical-layer security of offloaded data in a NOMA-MEC system, a max-min antieavesdropping poroblem was presented in~\cite{Rev1-3}.

Our system model is unique in a sense that it exploits NOMA to offload data of multiple users in a single time slot for mobile edge computations without impacting the performance of users. We solve the formulated max-min non-convex problem using the framework of successive convex approximation (SCA). A provably convergent algorithm is derived based on novel techniques to bound non-convex functions appearing in the main optimization problem. Precisely, our main contributions include: (i)~considering a novel model that integrates NOMA with MEC in a multi-user scenario, the main problem of maximizing the number of bits offloaded by the worst user to the MEC server using NOMA is formulated; (ii)~deriving a tractable and provably convergent algorithm using novel bounding techniques and based on successive convex optimization strategy to solve the formulated problem, and (iii)~numerical demonstration of the superiority of the proposed solution compared to the corresponding orthogonal multiple access (OMA) based approach.



\section{System Model and Problem Statement\label{SystemModel}}

\begin{figure}
\centering
\includegraphics[width=1\columnwidth]{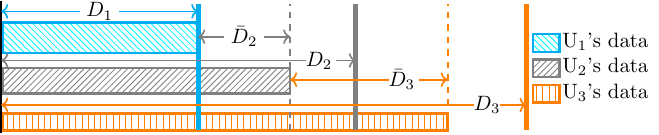}
\caption{Timing diagram of data offloading in a three user NOMA-MEC system. The duration shown by $D_1, D_2$ and $D_3$ denote the hard deadlines, and those shown by $\bar D_2$ and $\bar D_3$ denote the additional times.}
\label{fig:TimingDiagram}
\end{figure}

Consider a NOMA-MEC system consisting of a base station (BS) which is assumed to have an integrated MEC server, and a set of $M$ users, denoted by $\mathcal{U}\triangleq\left\{ \mathrm{U}_{1},\mathrm{U}_{2},\ldots,\mathrm{U}_{M}\right\} $. It is assumed that the BS and all users are each equipped with a single antenna. Furthermore, we assume that the users have very limited data processing capability, and therefore, they offload their computationally-intensive tasks to the MEC server. The channel gain between $\mathrm{U}_{m}$ ($m\in\mathcal{M}\triangleq\{1,2,\ldots,M\}$) and the BS is denoted by $g_{m}$. The deadline for $\mathrm{U}_{m}$ to offload its task to the MEC server is denoted by $D_{m}$, and without loss of generality, we consider $D_{1}\leq D_{2}\leq\cdots\leq D_{M}$.
It is also assumed that a user stops transmitting immediately even if it completely offloads its data before the assigned deadline. In particular, after $\mathrm{U}_{m-1}$ finishes offloading its data, $\mathrm{U}_{m}$, which also transmits in previous users' time slot, is allowed some additional time $\bar{D}_{m}$ to finish offloading its data until the hard deadline $D_{m}$, such that $\sum_{j=1}^m \bar D_j \leq D_m$.\footnote{To simplify the exposition, throughout the paper $\bar{D}_{1}$ is understood as $D_{1}$, if not mentioned otherwise.} As a result, users $m$ to $M$ simultaneously offload their data till $\bar{D}_{m}$ to the MEC server by means of NOMA. A timing diagram for an example of data offloading in a three-user NOMA-MEC system is shown in Fig.~\ref{fig:TimingDiagram}. For sum rate, an arbitrary decoding order can be used for successive interference cancellation (SIC) in uplink NOMA~\cite{DelayMinimization_Sherman,Rev2-5}. Nonetheless, optimal decoding order in the NOMA uplink with fairness consideration is relatively unexplored \cite{Rev2-4}. For the purpose of this work, we have fixed user decoding order according to their delay sensitivity, i.e., we consider SIC with respect to users' delay ordering such that higher-indexed users are decoded first, similar to the arguments in~\cite{Ding_MEC,Ding-MEC1,DelayMinimization_Sherman} and~\cite{Rev2-6}. In this way, the lower-indexed users (i.e., with stricter offloading deadlines) do not see interference from the higher indexed ones that transmit in their time slots. It should be emphasized here that NOMA is a natural candidate to integrate with MEC for the type of system model considered in our paper. Particularly, since users also transmit in previous users' time slot, NOMA, with its interference cancellation and power adaptation capabilities, is required for the smooth operation of users impacted by such interference.

The transmit power of $\mathrm U_m$ during $\bar{D}_ {j}$, $j \in \{1,2,\ldots,m\}$, is denoted by $P_{mj}$. Also, by properly normalizing the channel gains, we assume that the noise at the BS is distributed as zero-mean unit-variance circularly symmetric complex Gaussian random variable, $\mathcal{CN}\left(0,1\right)$. Therefore, letting $B$ denote the total available bandwidth, the data offloaded (in nats) by $\mathrm{U}_{m}$ to the MEC server is given by
\begin{equation}
\!N_{\text{off},m}=B\sum_{j=1}^{m}\bar{D}_{j}\ln\left(\!1\!+\!\frac{g_{m}P_{mj}}{\sum_{i=j}^{m-1}g_{i}P_{ij}+1}\!\right),m\in\mathcal{M}.\label{eq:rateexpression}
\end{equation}
Note that the expression for the rate of $\mathrm U_m$ (i.e., $\ln(1+\frac{g_{m}P_{mj}}{\sum_{i=j}^{m-1}g_{i}P_{ij}+1})$) in~\eqref{eq:rateexpression} satisfies the condition for (deadline-dependent) SIC ordering as discussed in~\cite{Ding_MEC} and~\cite{Ding-MEC1}. 

In this paper, we are interested to find the power allocation $P_{mj}$ and additional times $\bar{D}_{m}$ to maximize the worst task to be offloaded among all users, which is formulated as  
\begin{subequations}
\label{MEC:NOMA} 
\begin{align}
{\mathop{{\rm maximize}}\limits _{\{P_{mj}\geq0,\bar{D}_{j}\geq0\}}} & \ \underset{m \in \mathcal M}{\min}N_{\text{off},m},\label{eq:MainObj}\\
{\rm subject~to} & \ \sum\nolimits _{m=1}^{M}\sum\nolimits _{j=1}^{m}\bar{D}_{j}P_{mj}\leq E_{\textrm{th}},\label{eq:ener}\\
 & \ \sum\nolimits _{j=1}^{m}\bar{D}_{j}\leq D_{m}, m\in\mathcal{M}\setminus\{1\},\label{eq:deadline}\\
 & \!\!\!\!\!\!\!\! P_{m j}\leq P_{m}, j\in \mathcal M, m \in \{j, j+1, \ldots, M\}.\label{eq:txpower}
\end{align}
\end{subequations}
It is noteworthy that~\eqref{MEC:NOMA} is different from the conventional max-min rate optimization problem due to the presence of $\bar D_j$ in~\eqref{eq:rateexpression}, which is one of the optimization variables. The main motivation for considering this problem stems from promoting fairness among users to offload their data to the MEC server for computational purposes. If this consideration is ignored, the users that are more delay sensitive, and are unable to offload more data to meet delay constraints due to impeding channel conditions, suffer the most.\footnote{If resource allocation for all offloading users is to be considered on individual basis, the cost function of weighted sum of users' data to be offloaded may instead be considered. This problem is beyond the scope of current work and thus is left for future work.} We further remark that, to the best of our knowledge, \eqref{MEC:NOMA} has not been studied previously. In addition, the formulation of~\eqref{MEC:NOMA} is different from that of classical uplink NOMA systems~\cite{Rev3-2} both in terms of its objective and constraints. In~\eqref{eq:ener} we have constrained the total energy to $E_{\mathrm{th}}$,~\eqref{eq:deadline} ensures that $\textrm{U}_{m}$ completes its data offloading before its assigned deadline $D_{m}$ (with $\bar{D}_{1}=D_{1}$, which is not an optimization variable), and in~\eqref{eq:txpower} the  transmit power of $\mathrm U_{m}$ during any time slot $j$ with $j \in \mathcal M$, $m \in \{j, j+1, \ldots, M\}$ is constrained to $P_{m}$. It is also important to note that the constraint in~\eqref{eq:deadline} does not prevent a user with a longer deadline from offloading its data before a user with a shorter one. Specifically, such a scenario will occur when the interference encountered by a higher-indexed user in time slots of lower indexed users is small enough so that it is able to offload its data even before the lower indexed users have completed offloading their tasks. If such a scenario does not prevail, the higher indexed users are allowed to continue transmitting their data till the given task offloading deadline. We remark that unlike downlink NOMA, in an uplink NOMA based system an explicit SIC constraint is not required as each user is to be decoded only once by the BS~\cite{Rev4-1}. To keep the problem formulation more tractable, computation rates of user devices and MEC server have not been included as optimization variables in the problem formulation. Moreover, in line with the approximation used in existing works~\cite{Ding-MEC1,Ding_MEC},~\cite{Rev2-3,Rev2-6} we have ignored the processing energy and time to send data back to users from the MEC server in our problem statement. Inclusion of the mentioned above parameters is a ripe direction for future research. 

\section{Proposed Solution\label{PropSol}}
In this section, we propose an efficient solution to problem~\eqref{MEC:NOMA} and prove the convergence of the proposed algorithm. Moreover, for a fair comparison, we also discuss a corresponding optimized OMA-MEC system. Since the problem in \eqref{MEC:NOMA} is non-convex, to solve it we  approximate the non-convex constraints with convex ones. Afterwards, we use the SCA framework to develop an iterative procedure to solve the original problem. One of novel features of this work lies in how we develop approximating functions that satisfy certain conditions within the SCA framework, as we see below.

For ease of description, we define $\mathbf{p}_{mj}\triangleq[P_{jj},P_{(j+1)j},\ldots,P_{mj}]\trans$ for $m\geq j$, which consists of the power allocation coefficients of $\mathrm{U}_{j}$ to $\mathrm{U}_{m}$ during $\bar{D}_{j}$, and $\mathbf{d}\triangleq[\bar{D}_{2},\bar{D}_{3},\ldots,\bar{D}_{M}]\trans$. It is obvious that the non-convexity of~\eqref{MEC:NOMA} is due to that of the objective in~\eqref{eq:MainObj}, and the constraint in~\eqref{eq:ener}. Let us consider an equivalent reformulation of \eqref{MEC:NOMA} given by \begin{subequations}
\label{MEC:NOMA:epigraph} 
\begin{align}
\mathop{{\rm maximize}}\limits _{\mathbf{p}\geq0,\mathbf{d}\geq 0,\ell} & \quad\ell\\
{\rm subject~to} & \quad N_{\text{off},m}\geq\ell,m\in\mathcal{M},\label{eq:rate}\\
 & \quad\eqref{eq:ener},\eqref{eq:deadline},\eqref{eq:txpower}
\end{align}
where 
\[
\mathbf{p}=[\mathbf{p}_{MM}\trans,\ldots,\mathbf{p}_{M1}\trans,\mathbf{p}_{(M-1)(M-1)}\trans,\ldots,\mathbf{p}_{(M-1)1}\trans,\ldots,\mathbf{p}_{11}\trans]\trans.
\]
\end{subequations}
It is clear that the non-convexity of~\eqref{MEC:NOMA} is carried over into ~\eqref{eq:rate} and~\eqref{eq:ener}. Let us focus on~\eqref{eq:rate} first. In light of SCA method, to deal with~\eqref{eq:rate} we need to find \emph{a concave lower bound} of $N_{\text{off},m}$. In the related literature, a popular method is to introduce some auxiliary variables to make~\eqref{eq:rate} more tractable, which eventually increases the complexity of the convex subproblems. Instead, to approximate~\eqref{eq:rate} we present the following theorem.
\begin{thm}
\label{thm:1stApp}For given $x_{0}\geq0$, $u_{0}\geq0$, and $v_{0}>0$, the inequality in~\eqref{eq:LB}, shown on the next page, holds for any $u\geq0$ and $v>0$. 
\begin{figure*}
\begin{centering}
\begin{small}
\begin{align}
 & x\ln\bigg(1+\dfrac{u}{v}\bigg)\geq\dfrac{1-x_{0}}{2}-\dfrac{(x_{0}-v_{0})^{2}}{4v_{0}}+\bigg[2+\ln\bigg(1+\dfrac{u_{0}}{v_{0}}\bigg)+\dfrac{x_{0}-1}{2}+\dfrac{1}{2}\bigg(\dfrac{x_{0}}{v_{0}}-1\bigg)\bigg]x-\bigg[\dfrac{(x_{0}-1)^{2}}{4(u_{0}+v_{0})}+\dfrac{1}{2}\bigg(\dfrac{x_{0}}{v_{0}}-1\bigg)\bigg]v\nonumber \\
 & \qquad\qquad\qquad\qquad\qquad\qquad\qquad\qquad\qquad\qquad\qquad\qquad\qquad\qquad-\dfrac{(x_{0}-1)^{2}}{4(u_{0}+v_{0})}u-\dfrac{(x+v)^{2}}{4v_{0}}-\dfrac{(u_{0}+v_{0})(x+1)^{2}}{4(u+v)}.\label{eq:LB}
\end{align}
\end{small}
\par\end{centering}
\vskip-0.2in
\end{figure*}
\end{thm}
\begin{IEEEproof}
See Appendix~\ref{Proof}. 
\end{IEEEproof}
We also remark that the derivative of both sides of~\eqref{eq:LB} with respect to each variable involved is identical, which is another requirement for the use of a lower bound in light of the SCA principles.

Let $\mathcal{I}_{mj}\triangleq\sum_{i=j}^{m-1}g_{i}P_{ij}+1$, which is an affine function of $\{P_{ij}\}_{i=j}^{m-1}$. Then \eqref{eq:rateexpression} is equivalent to
\begin{equation}
N_{\text{off},m}=B\sum \nolimits_{j=1}^{m}\bar{D}_{j}\ln\Bigl(1+g_{m}P_{mj}/\mathcal{I}_{mj}\Bigr).\label{NOMA:Data_Off}
\end{equation}
Now, let $P_{mj}^{(n)}$ and $\bar{D}_{j}^{(n)}$ be the values of $P_{mj}$ and $\bar{D}_{j}$ in iteration $n$ of the iterative process presented in the discussion to follow, respectively. Then, applying Theorem \ref{thm:1stApp} we can approximate \eqref{eq:rate} as 
\begin{equation}
B\sum \nolimits_{j=1}^{m}f_{mj}\bigl(\mathbf{p}_{mj},\bar{D}_{j};\mathbf{p}_{mj}^{(n)},\bar{D}_{j}^{(n)}\bigr)\geq\ell,m\in\mathcal{M},\label{eq:rate:app}
\end{equation}
where $f_{m,j}\bigl(\mathbf{p}_{mj},\bar{D}_{j};\mathbf{p}_{mj}^{(n)},\bar{D}_{j}^{(n)}\bigr)$
is given by~\eqref{eq:fDef} (shown on the next page), $\mathcal{I}_{mj}^{(n)}\triangleq\sum_{i=j}^{m-1}P_{ij}^{(n)}g_{i}+1,$
$a_{mj}^{(n)}\triangleq0.5\big(1-\bar{D}_{j}^{(n)}\big)-0.25\big(\bar{D}_{j}^{(n)}-\mathcal{I}_{mj}^{(n)}\big)^{2}/\mathcal{I}_{mj}^{(n)},$
$b_{mj}^{(n)}\triangleq2+\ln\big(1+g_{m}P_{mj}^{(n)}/\mathcal{I}_{mj}^{(n)}\big)+0.5\big(\bar{D}_{j}^{(n)}-1\big)+0.5\big(-1+\bar{D}_{j}^{(n)}/\mathcal{I}_{mj}^{(n)}\big),$
$c_{mj}^{(n)}\triangleq0.25\big(\bar{D}_{j}^{(n)}\!-1\big)^{2}/\big(g_{m}P_{mj}^{(n)}+\mathcal{I}_{mj}^{(n)}\big)+0.5\big(-1+\bar{D}_{j}^{(n)}/\mathcal{I}_{mj}^{(n)}\big),$
$d_{mj}^{(n)}\triangleq0.25\big(\bar{D}_{j}^{(n)}\!-1\big)^{2}g_{m}/\big(g_{m}P_{mj}^{(n)}+\mathcal{I}_{mj}^{(n)}\big),$
and $e_{mj}^{(n)}\triangleq0.25\big(g_{m}P_{mj}^{(n)}+\mathcal{I}_{mj}^{(n)}\big)$.

\begin{figure*}
\begin{small}
\begin{equation}
f_{m,j}\big(\mathbf{p}_{mj},\bar{D}_{j};\mathbf{p}_{mj}^{(n)},\bar{D}_{j}^{(n)}\bigr)=a_{mj}^{(n)}+b_{mj}^{(n)}\bar{D}_{j}-c_{mj}^{(n)}\mathcal{I}_{mj}-d_{mj}^{(n)}P_{mj}-\dfrac{(\bar{D}_{j}+\mathcal{I}_{mj})^{2}}{4\mathcal{I}_{mj}^{(n)}}-e_{mj}^{(n)}\dfrac{(\bar{D}_{j}+1)^{2}}{(g_{m}P_{mj}+\mathcal{I}_{mj})}.\label{eq:fDef}
\end{equation}
\end{small}
\hrulefill
\end{figure*}
\noindent We remark that~\eqref{eq:rate:app} can be expressed as a few simple second-order cone (SOC) constraints. To this end, auxiliary variables are introduced to obtain a system of linear and quadratic inequalities such that the quadratic inequalities are transformed to SOC constraints using simple manipulations~\cite{Nam:WSRmax:2012}. This is accomplished automatically by a modeling tool such as CVX.

Our next step is to approximate~\eqref{eq:ener} which requires a convex upper bound of the term $\bar{D}_{j}P_{mj}$. There are several ways to achieve this. For example, the authors in~\cite{Nam:WSRmax:2012} used $\bar{D}_{j}P_{mj}\leq\frac{\phi}{2}\bar{D}_{j}^{2}+\frac{1}{2\phi}P_{mj}^{2}$, which holds good for any $\phi>0$, and the equality is achieved when $\phi=P_{mj}/\bar{D}_{j}$. However, in this paper $\bar{D}_{j}$ and/or $P_{mj}$ \emph{can approach zero}, which may lead to some numerical issues if the above convex upper bound is used. Thus, in this paper we simply use the following convex upper bound
\begin{align}
& \!\!\!\!\bar{D}_{j}P_{mj} \leq0.25\bigl(\bar{D}_{j}+P_{mj}\bigr)^{2}+0.25\bigl(\bar{D}_{j}^{(n)}-P_{mj}^{(n)}\bigr)^{2} -0.5 \nonumber \\
 & \!\!\!\!\!\!\times \bigl(\bar{D}_{j}^{(n)}\!-\!P_{mj}^{(n)}\bigr)\bigl(\bar{D}_{j}\!-\!P_{mj}\bigr) \triangleq \! q_{mj}(P_{mj},\bar{D}_{j};P_{mj}^{(n)},\bar{D}_{j}^{(n)}),
\end{align}
which approximates~\eqref{eq:ener} as 
\begin{equation}
\sum\nolimits _{m=1}^{M}\sum\nolimits _{j=1}^{m}q_{mj}(P_{mj},\bar{D}_{j};P_{mj}^{(n)},\bar{D}_{j}^{(n)})\leq E_{\textrm{th}}.\label{eq:ener:app}
\end{equation}
We remark that $q_{mj}(P_{mj},\bar{D}_{j};P_{mj}^{(n)},\bar{D}_{j}^{(n)})$ is in fact symbolic representation of a convex quadratic function and thus \eqref{eq:ener:app} is an SOC constraint. From the convex approximation of \eqref{eq:rate} and \eqref{eq:ener} presented above, we arrive at the following second-order cone program (SOCP):
\begin{subequations}
\label{MEC:NOMA:subprob}
\begin{align}
\mathop{{\rm maximize}}\limits _{\mathbf{p}\geq0,\mathbf{d}\geq 0,\ell} & \quad\ell\\
{\rm subject~to} & \quad\eqref{eq:deadline},\eqref{eq:txpower},\eqref{eq:rate:app},\eqref{eq:ener:app}.\label{eq:rate-1}
\end{align}
\end{subequations}
 Now we are in a position to present an algorithmic framework to solve our main problem in~\eqref{MEC:NOMA}, using the approximation developed in~\eqref{MEC:NOMA:subprob}. The procedure is outlined in \textbf{Algorithm~\ref{alg:DORM}}. We simply set $P_{mj}^{(0)}=0$ and $\bar{D}_{j}^{(0)}=0$ as a starting feasible point.  
\begin{algorithm}[t]
\caption{Offloading rate maximization algorithm}
\label{alg:DORM}
\KwIn{ $P_{mj}^{(0)}$, $\bar{D}_{j}^{(0)}$;$\forall m,j\in\mathcal{M}$}
\KwOut{ $\text{\ensuremath{P_{mj}^{\star}}},\text{\ensuremath{D_{j}^{\star}}};\text{\ensuremath{\forall m,j\in\mathcal{M}}}$}
$n\leftarrow0$\;
\Repeat{convergence }{
Solve~\eqref{MEC:NOMA:subprob} and denote the obtained solution
as $P_{mj}^{\star}$ and $D_{j}^{\star}$\;
Update: $P_{mj}^{(n+1)}\leftarrow P_{mj}^{\star},$ $\bar{D}_{j}^{(n+1)}\leftarrow D_{j}^{\star}$\;
$n\leftarrow n+1$
}
\end{algorithm}
It is quite insightful to observe that for each run of the proposed algorithm, we solve a convex problem whose feasible set is a subset of the original problem in~\eqref{MEC:NOMA}. This occurs due to the fact that we approximate the non-convex function in~\eqref{NOMA:Data_Off} using a lower bound as given in~\eqref{eq:rate:app}. It not only ensures that solution in each iteration of~\textbf{Algorithm~\ref{alg:DORM}} is feasible to~\eqref{MEC:NOMA}, but also results in convergence as we see in the next section.
\subsection{Convergence of \textbf{Algorithm \ref{alg:DORM}}}
Let the feasible set of~\eqref{MEC:NOMA:epigraph} be denoted by $\mathcal{F}$; $\mathcal{F}^{(n)}\triangleq\{(\mathbf{p},\mathbf{d},\ell)\ \textrm{satisfying constraints in \eqref{MEC:NOMA:subprob}}\}$, and $\ell^{(n)}$ denote the feasible set and the optimal objective at the $n^{\textrm{th}}$ iteration, respectively. Let the \emph{non-convex constraints} in \eqref{MEC:NOMA:epigraph} be compactly denoted by $\psi_{i}(\mathbf{p},\mathbf{d},\ell)$, and the corresponding approximated convex constraints in \eqref{MEC:NOMA:subprob} by $\Psi_i(\mathbf{p},\mathbf{d},\ell;\mathbf{p}^{(n)},\mathbf{d}^{(n)},\ell^{(n)})$. It is straightforward to note that due to the proposed bounds, the convex constraints used to approximate the non-convex ones satisfy the following three properties 
\begin{subequations}
\begin{align}
 &\Psi_i\bigl(\mathbf{p},\mathbf{d},\ell;\mathbf{p}^{(n)},\mathbf{d}^{(n)},\ell^{(n)}\bigr)\geq \psi_i\bigl(\mathbf{p},\mathbf{d},\ell\bigr),\forall\mathbf{p},\mathbf{d},\ell\label{propA}\\
 & \Psi_i(\mathbf{p}^{(n)},\mathbf{d}^{(n)},\ell^{(n)};\mathbf{p}^{(n)},\mathbf{d}^{(n)},\ell^{(n)})=\psi_i\bigl(\mathbf{p}^{(n)},\mathbf{d}^{(n)},\ell^{(n)}\bigr)\label{propB}\\
 & \nabla \Psi_i(\mathbf{p}^{(n)},\mathbf{d}^{(n)},\ell^{(n)};\mathbf{p}^{(n)},\mathbf{d}^{(n)},\ell^{(n)})=\nabla \psi_i\bigl(\mathbf{p}^{(n)},\mathbf{d}^{(n)},\ell^{(n)}\bigr).\label{propC}
\end{align}
\end{subequations}
First of all we note that $\mathcal{F}^{(n)}$ is a subset of the feasible set of the original problem due to \eqref{propA}. Further, as a consequence of \eqref{propB} and the update rule given in \textbf{Algorithm~\ref{alg:DORM}}, the sequence $\bigl(\mathbf{p}^{(n)},\mathbf{d}^{(n)},\ell^{(n)}\bigr)$ is feasible to both $\mathcal{F}^{(n)}$ and $\mathcal{F}^{(n+1)}$. Since $\bigl(\mathbf{p}^{(n)},\mathbf{d}^{(n)},\ell^{(n)}\bigr)$ is feasible to $\mathcal{F}^{(n+1)}$, the optimal variables corresponding to $\mathcal{F}^{(n+1)}$ should yield an improved objective value in the $(n+1)^{\textrm{th}}$ run of the algorithm i.e., $\ell^{(n)}\leq\ell^{(n+1)}$. It is easy to see that $\mathcal{F}$ is a compact set, and thus, the sequence \{$\ell^{(n)}$\} converges.

We also remark that Slater's condition holds for the subproblem in all iterations. This can be shown by choosing sufficiently small $(\mathbf{p},\mathbf{d},\ell)$. Thus, the Karush-Kuhn-Tucker (KKT) conditions must be satisfied for \eqref{MEC:NOMA:subprob} in all iterations. Let $\bigl(\mathbf{p}^{\ast},\mathbf{d}^{\ast},\ell^{\ast}\bigr)$ be the limit of the sequence $\bigl(\mathbf{p}^{(n)}, \mathbf{d}^{(n)},\ell^{(n)}\bigr)$.\footnote{Following \cite{Inosha_Topology}, a quadratic proximal term can be added to the non-strict convex objective in \eqref{MEC:NOMA:subprob}, without effecting the optimal point, so as to ensure the existence of such limit of the sequence.} Without loss of generality we assume that $\bigl(\mathbf{p}^{(n)},\mathbf{d}^{(n)},\ell^{(n)}\bigr)\to\bigl(\mathbf{p}^{\ast},\mathbf{d}^{\ast},\ell^{\ast}\bigr)$. Otherwise we can restrict to a subsequence. Thus it holds that $\Psi_i(\mathbf{p}^{(n+1)},\mathbf{d}^{(n+1)},\ell^{(n+1)};\mathbf{p}^{(n)},\mathbf{d}^{(n)},\ell^{(n)})\to\Psi_i(\mathbf{p}^{\ast},\mathbf{d}^{\ast},\ell^{\ast};\mathbf{p}^{\ast},\mathbf{d}^{\ast},\ell^{\ast})\!=\!\psi_i\bigl(\mathbf{p}^{\ast},\mathbf{d}^{\ast},\ell^{\ast}\bigr)$.
Also, due to \eqref{propC}, we have $\nabla\Psi_i\bigl(\mathbf{p}^{(n+1)},\mathbf{d}^{(n+1)},\ell^{(n+1)};\mathbf{p}^{(n)},\mathbf{d}^{(n)},\ell^{(n)}\bigr)\to\nabla\Psi_i\bigl(\mathbf{p}^{\ast},\mathbf{d}^{\ast},\ell^{\ast};\mathbf{p}^{\ast},\mathbf{d}^{\ast},\ell^{\ast}\bigr)=\nabla \psi_i\bigl(\mathbf{p}^{\ast},\mathbf{d}^{\ast},\ell^{\ast}\bigr)$. The same arguments also apply to the convex constraints in \eqref{MEC:NOMA}. Thus if $\bigl(\mathbf{p}^{\ast},\mathbf{d}^{\ast},\ell^{\ast}\bigr)$ is a regular point of \eqref{MEC:NOMA:epigraph}, then we can show that $\bigl(\mathbf{p}^{\ast},\mathbf{d}^{\ast},\ell^{\ast}\bigr)$ is also a KKT point of \eqref{MEC:NOMA:epigraph}. We skip the details here for the sake of brevity.
\begin{figure*}[t]
\centering
	\begin{minipage}{.24\textwidth}
		\centering
		\pgfplotsset{compat=1.9}
\resizebox{0.95\columnwidth}{0.73\columnwidth}{
\begin{tikzpicture}
  \begin{axis}[
      	xlabel={\Large{Iteration number}},
	xtick={0, 5, 10, 15, 20},
	xmin=0,xmax=20,
	ylabel={\Large{Avg. min. \# offloaded bits (kb)}},
	ytick={0, 50, 100, 150, 200, 250, 300,350},
	ymin=0,ymax=350,
	grid=both,
	minor grid style={gray!25},
	major grid style={gray!25},
	legend columns=1, 
legend style={{nodes={scale=1.15, transform shape}}, at={(0.53,0.19)},  anchor=south west, draw=none,fill=white,legend cell align=left,inner sep=1pt,row sep = -2pt}]
	\addplot[line width=1pt,solid,color=cyan,mark=*,mark options={solid,cyan},mark size=2.5pt] table [y=NOMA_5_5, x=Iteration,col sep = comma]{Convergence.csv};
	\addlegendentry{NOMA (5, 5)}
	\addplot[line width=1pt,solid,color=orange,mark=triangle*,mark options={solid,orange},mark size=2.5pt] table [y=NOMA_4_5, x=Iteration,col sep = comma]{Convergence.csv};
	\addlegendentry{NOMA (4, 5)}
	\addplot[line width=1pt,solid,color=gray,mark=diamond*,mark options={solid,gray},mark size=2.5pt] table [y=NOMA_5_4, x=Iteration,col sep = comma]{Convergence.csv};
	\addlegendentry{NOMA (5, 4)}
	\addplot[line width=1pt,dotted,color=cyan,mark=o,mark options={solid,cyan},mark size=2.5pt] table [y=OMA_5_5, x=Iteration,col sep = comma]{Convergence.csv};
	\addlegendentry{OMA (5, 5)}
	\addplot[line width=1pt,dashed,color=orange,mark=triangle,mark options={solid,orange},mark size=2pt] table [y=OMA_4_5, x=Iteration,col sep = comma]{Convergence.csv};
	\addlegendentry{OMA (4, 5)}
	\addplot[line width=1pt,dashed,color=gray,mark=diamond,mark options={solid,gray},mark size=2.5pt] table [y=OMA_5_4, x=Iteration,col sep = comma]{Convergence.csv};
	\addlegendentry{OMA (5, 4)}

	\end{axis}
\end{tikzpicture}
}
		\vskip-0.1in
		\caption{Convergence of the NOMA-MEC and OMA-MEC systems.}
		\label{fig:Convergence}
	\end{minipage}
	\hfill
	\begin{minipage}{.24\textwidth}
		\centering
		\pgfplotsset{compat=1.9}
\resizebox{0.95\columnwidth}{0.73\columnwidth}{
\begin{tikzpicture}
  \begin{axis}[
      	xlabel={\Large{$E_{\mathrm{th}}$ (mJ)}},
	xtick={1, 2, 3, 4, 5, 6, 7, 8, 9, 10},
	xmin=1,xmax=10,
	ylabel={\Large{Avg. min. \# offloaded bits (kb)}},
	ytick={0, 100, 200, 300, 400, 500, 600, 700, 800},
	ymin=0,ymax=800,
	grid=both,
	minor grid style={gray!25},
	major grid style={gray!25},
	legend columns=1, 
legend style={{nodes={scale=1.1, transform shape}}, at={(0,0.54)},  anchor=south west, draw=none,fill=white,legend cell align=left,inner sep=1pt,row sep = -2pt}]
	\addplot[line width=1.5pt,solid,color=cyan,mark=*,mark options={solid,cyan},mark size=3pt] table [y=NOMA_10, x=Eth,col sep = comma]{l_vs_Eth.csv};
	\addlegendentry{$P_\mathrm{t} = 10$ dBm}
	\addplot[line width=1.5pt,solid,color=gray,mark=triangle*,mark options={solid,gray},mark size=3.5pt] table [y=NOMA_8, x=Eth,col sep = comma]{l_vs_Eth.csv};
	\addlegendentry{$P_\mathrm{t} = 8$ dBm}
	\addplot[line width=1.5pt,solid,color=orange,mark=diamond*,mark options={solid,orange},mark size=3.5pt] table [y=NOMA_5, x=Eth,col sep = comma]{l_vs_Eth.csv};
	\addlegendentry{$P_\mathrm{t} = 5$ dBm}
	\addplot[line width=1.5pt,dashed,color=cyan,mark=o,mark options={solid,cyan},mark size=3pt] table [y=OMA_10, x=Eth,col sep = comma]{l_vs_Eth.csv};
	\addlegendentry{$P_\mathrm{t} = 10$ dBm}
	\addplot[line width=1.5pt,dashed,color=gray,mark=triangle,mark options={solid,gray},mark size=3.5pt] table [y=OMA_8, x=Eth,col sep = comma]{l_vs_Eth.csv};
	\addlegendentry{$P_\mathrm{t} = 8$ dBm}
	\addplot[line width=1.5pt,dashed,color=orange,mark=diamond,mark options={solid,orange},mark size=3.5pt] table [y=OMA_5, x=Eth,col sep = comma]{l_vs_Eth.csv};
	\addlegendentry{$P_\mathrm{t} = 5$ dBm}	
	\node[align=center,fill=white,inner sep=3pt] (D1) at (axis cs: 6, 750) {\Large{NOMA}};
\draw[line width = 1.5pt] (axis cs:8, 650) arc (140:-135: 0.2cm and  1.60cm);
\draw[->,>=latex,line width = 1.5pt] (D1) -- (axis cs:8, 690);
	\node[align=center,fill=white,inner sep=3pt] (D1) at (axis cs: 6, 250) {\Large{OMA}};
\draw[line width = 1.5pt] (axis cs:7.8, 150) arc (140:-135: 0.2cm and  0.60cm);
\draw[->,>=latex,line width = 1.5pt] (D1) -- (axis cs:7.8, 180);
	\end{axis}
\end{tikzpicture}
}
		\vskip-0.1in
		\caption{Average minimum number of offloaded bits versus $E_{\mathrm{th}}$.}
		\label{fig:l_vs_Eth}
	\end{minipage}
	\hfill
	\begin{minipage}{.24\textwidth}
		\centering
		\pgfplotsset{compat=1.9}
\usetikzlibrary{patterns}
\resizebox{0.95\columnwidth}{0.73\columnwidth}{
\pgfplotstableread[col sep=comma]{l_vs_Users.csv}{\loadedtable}
\pgfplotstablegetcolsof{\loadedtable}
\pgfmathtruncatemacro{\NoOfCols}{\pgfplotsretval-1}
\begin{tikzpicture}   
	\begin{axis}
		[
  		ymin=0,
  		ymax=620,
  		ybar=0pt,
  		ytick={0, 200, 400, 620},
  		xtick align=inside,
  		ylabel={\Large{Avg. min. \# offloaded bits (kb)}},
  		xtick={3,4,5,6},
			xmin=3,xmax=6,
			xlabel={\Large{Number of users, $M$}},
      bar width=0.85/(\NoOfCols),  		
      enlarge x limits={abs=0.5},
      grid=both,
			minor grid style={gray!25},
			major grid style={gray!25},
			legend columns=1, 
legend style={{nodes={scale=1.2, transform shape}}, at={(0.55,0.495)},  anchor=south west, draw=none,fill=white,legend cell align=left,inner sep=1pt,row sep = -2pt}]
		\addplot[cyan,postaction={pattern=north east lines, pattern color=cyan}] table [x =M,y=NOMA_5_5,col sep=comma,] {\loadedtable};	
		\addlegendentry{NOMA (5, 5)}
		\addplot[cyan,fill=cyan] table [x =M,y=OMA_5_5,col sep=comma,] {\loadedtable};
		\addlegendentry{OMA (5, 5)}	
		\addplot[gray,postaction={pattern=north east lines,pattern color=gray}] table [x =M,y=NOMA_4_5,col sep=comma,] {\loadedtable};	
		\addlegendentry{NOMA (4, 5)}
		\addplot[gray,fill=gray] table [x =M,y=OMA_4_5,col sep=comma,] {\loadedtable};	
		\addlegendentry{OMA (4, 5)}
		\addplot[orange,postaction={pattern=north east lines,pattern color=orange}] table [x =M,y=NOMA_5_4,col sep=comma,] {\loadedtable};	
		\addlegendentry{NOMA (5, 4)}
		\addplot[orange,fill=orange] table [x =M,y=OMA_5_4,col sep=comma,] {\loadedtable};	
		\addlegendentry{OMA (5, 4)}
	\end{axis}      
\end{tikzpicture}
}
		\vskip-0.1in
		\caption{Average minimum number of offloaded bits versus $M$.}
		\label{fig:l_vs_M}
	\end{minipage} 
	\hfill
	\begin{minipage}{.24\textwidth}
		\centering
		\pgfplotsset{compat=1.9}
\resizebox{0.95\columnwidth}{0.73\columnwidth}{
\begin{tikzpicture}   
      \begin{axis}[
    xlabel={\Large{$\Delta$ (centiseconds)}},
	xtick={0.01, 0.02, 0.03, 0.04, 0.05, 0.06, 0.07, 0.08, 0.09, 0.1},
	xmin=0.01,xmax=0.1,
	xticklabels={1,2,3,4,5,6,7,8,9,10},
	ylabel={\Large{Avg. min. \# offloaded bits (kb)}},
	ytick={0, 100, 200, 300, 365},
	ymin=9,ymax=370,
	grid=both,
	minor grid style={gray!25},
	major grid style={gray!25},
	legend columns=1, 
legend style={{nodes={scale=1.1, transform shape}}, at={(0.45,0.25)},  anchor=south west, draw=none,fill=white,legend cell align=left,inner sep=1pt,row sep = -2pt}]
	\addplot[line width=1.5pt,solid,color=cyan,mark=triangle*,mark size=3.5pt] table [y=NOMA_5_5_0.5, x=Delta,col sep = comma]{Deadline.csv};
	\addlegendentry{NOMA $(5, 5, 0.5)$}
	\addplot[line width=1.5pt,solid,color=orange,mark=*,mark size=3pt] table [y=NOMA_5_5_0.3, x=Delta,col sep = comma]{Deadline.csv};
	\addlegendentry{NOMA $(5, 5, 0.3)$}
	\addplot[line width=1.5pt,dashed,color=cyan,mark=triangle,mark options={solid}, mark size=3.5pt] table [y=OMA_5_5_0.5, x=Delta,col sep = comma]{Deadline.csv};;
	\addlegendentry{OMA $(5, 5, 0.5)$}
	\addplot[line width=1.5pt,dashed,color=orange,mark=o,mark options={solid},mark size=3pt] table [y=OMA_5_5_0.3, x=Delta,col sep = comma]{Deadline.csv};
	\addlegendentry{OMA $(5, 5, 0.3)$}	

	\end{axis}
\end{tikzpicture}
}
		\vskip-0.1in
		\caption{Average minimum number of offloaded bits versus $\Delta$.}
		\label{fig:Deadline}
	\end{minipage} 
\end{figure*}
\subsection{Complexity of \textbf{Algorithm~\ref{alg:DORM}}}
As shown in Section \ref{Res}, \textbf{Algorithm \ref{alg:DORM}} converges in  few  iterations and in each iteration the SOCP in \eqref{MEC:NOMA:subprob} needs to be solved. The worst case complexity of generic interior-point methods for an SOCP depends on the number of optimization variables, number of SOC constraints and the dimension of SOC constraints \cite{Boyd_SOCP}. In our case, \eqref{MEC:NOMA:subprob} has $(M^2+2M)$ optimization variables, $O(M^2)$ SOC constraints and the total dimension of SOC constraints\footnote{Big-O notation has been used in literature to represent the number of variables, SOC constraints and their dimension \cite{Rev1-Comp}.} is $O(M^2)$. Using the complexity estimate of \cite{Boyd_SOCP}, the accuracy of the duality gap can be improved by a constant factor in $O((M^2)^{1/2})$ iterations and the time complexity of each iteration is $O(M^4\times M^2)$. This leads to an overall complexity of $O(M^7)$.
This complexity bound is obtained by simply viewing each linear inequality constraint of the form $\mathbf{c}\trans \mathbf{x} +d \geq 0 $ 
as a special case of the SOC constraint $\mathbf{c}\trans \mathbf{x} +d \geq  \lVert\mathbf{A}\mathbf{x}+\mathbf{b}\rVert_2$, where $\mathbf{A}=0$ and $\mathbf{b}=0$. Consequently, the problem data of the resulting SOCP consists of many zeros. Modern off-the-shelf conic solvers can efficiently exploit this sparsity, and thus the practical run time for solving \eqref{MEC:NOMA:subprob} is much less than  the worst-case time complexity. A further improvement in computational complexity can be obtained by using first-order optimization methods to solve the given problem in future study.
\subsection{A corresponding OMA-MEC system}
In order to showcase the advantages of the NOMA-based system, in this subsection, we present a corresponding baseline OMA-based system. Following the two-user (pure) OMA system in~\cite{Ding-MEC1}, we consider a multi-user OMA-MEC system, where a dedicated time slot is allocated to every user. Note that we reuse the notations from the NOMA-MEC system in this subsection when appropriate. Specifically, $\mathrm{U}_{m}$ offloads its data to the MEC server \textit{only} during $\hat{D}_{m}$ such that $\sum_{j=1}^{m}\hat{D}_{j}\leq D_{m}$. The data offloaded (in nats) by $\mathrm{U}_{m}$ to the server is given by $\hat{N}_{\mathrm{off},m}=B\hat{D}_{m}\ln\left(1+g_{m}\hat{P}_{m}\right),$ where $\hat{P}_{m}$ is the power transmitted by $\mathrm{U}_{m}$ during $\hat{D}_{m}.$ The problem of maximizing the minimum number of offloaded nats among all users in the OMA-MEC can be easily formulated but is skipped here due to the space constraint. Similar to the case of NOMA, we can derive a SCA-based method to solve the resulting problem. Explicitly in iteration $n+1$, an SOCP can be formulated as follows: 
\begin{subequations}
\label{OptProb_OMA}
\begin{align}
\underset{\hat{\mathbf{p}}\geq0,\hat{\mathbf{d}}\geq0}{\!\!\!\!\mathrm{maximize}}\  & \varphi,\label{eq:OMA_Slack}\\
\!\!\!\!\mathrm{\!\!\!\!subject}\ \mathrm{to}\  & B \hat{f}\big(\hat{P}_{m},\hat{D}_{m};\hat{P}_{m}^{(n)},\hat{D}_{m}^{(n)}\big)\geq\varphi,\label{eq:OMA_ObjLargerSlack}\\
 & \!\!\!\!\!\!\sum_{m=1}^{M}\hat{q}_{m}(\hat{P}_{m},\hat{D}_{m};\hat{P}_{m}^{(n)},\hat{D}_{m}^{(n)})\leq E_{\mathrm{th}},m\in\mathcal{M},\label{eq:OMA_EnergyConstraint}\\
 & \!\!\!\!\!\!\sum \nolimits_{j=1}^{m}\hat{D}_{j}\leq D_{m},m\in\mathcal{M},\label{eq:OMA_DelayConstraint}\\
 & \!\!\!\!\!\!\hat{P}_{m}\leq P_{m},m\in\mathcal{M},\label{eq:OMA_PowerConstraint}
\end{align}
\end{subequations}
where $\hat{\mathbf{p}}\triangleq[\hat{P}_{1},\hat{P}_{2},\ldots,\hat{P}_{M}]\trans$,
$\hat{\mathbf{d}}\triangleq[\hat{D}_{1},\hat{D}_{2},\ldots,\hat{D}_{M}]\trans$,
and $B\hat{f}_{m}\big(\hat{P}_{m},\hat{D}_{m};\hat{P}_{m}^{(n)},\hat{D}_{m}^{(n)}\big)$
is a concave lower bound on $\hat{N}_{\mathrm{off},m}$,
$\hat{f}_{m}\big(\hat{P}_{m},\hat{D}_{m};\hat{P}_{m}^{(n)},\hat{D}_{m}^{(n)}\big)\triangleq\hat{a}_{m}+\hat{b}_{m}\hat{D}_{m}-\hat{c}_{m}\hat{P}_{m}-\hat{d}_{m}(\hat{D}_{m}+1)^{2}/(1+g_{m}\hat{P}_{m})$,
$\hat{a}_{m}\triangleq0.25\left[1-\big(\hat{D}_{m}^{(n)}\big)^{2}+\big(\hat{D}_{m}^{(n)}-1\big)^{2}g_{m}\hat{P}_{m}^{(n)}/\big(1+g_{m}\hat{P}_{m}^{(n)}\big)\right],$
$\hat{b}_{m}\triangleq\big[1+\ln(1+g_{m}\hat{P}_{m}^{(n)})+0.5\big(\hat{D}_{m}^{(n)}-1\big)\big],$
$\hat{c}_{m}\triangleq0.25g_{m}\big(\hat{D}_{m}^{(n)}-1\big)^{2}/\big(1+g_{m}\hat{P}_{m}^{(n)}\big)$
and $\hat{d}_{m}\triangleq0.25\big(1+g_{m}\hat{P}_{m}^{n)}\big).$
Similarly, we have $\hat{q}_{m}(\hat{P}_{m},\hat{D}_{m};\hat{P}_{m}^{(n)},\hat{D}_{m}^{(n)})\triangleq0.25(\hat{D}_{m}+\hat{P}_{m})^{2}+0.25(\hat{D}_{m}^{(n)}-\hat{P}_{m}^{(n)})^{2}-0.5(\hat{D}_{m}^{(n)}-\hat{P}_{m}^{(n)})(\hat{D}_{m}-\hat{P}_{m})\geq\hat{D}_{m}\hat{P}_{m}.$\\
Although a dedicated interference free time slot is allocated to each user in OMA system to offload data to MEC server, yet a combination of power controlled data offloading of users to previous time slots and low interference at delay sensitive users due to SIC make NOMA-MEC perform better for MEC purposes. In the section to follow, we perform various numerical experiments that corroborate this observation.

\begin{table}[t]
\centering
\caption{Average problem solving time for NOMA-MEC system using~\textbf{Algorithm~\ref{alg:DORM}}.}
\label{TimeTable}
\begin{tabular}{|l|l|l|l|l|l|}
\hline
\# of users, $M$  & $4$ & $8$ & $12$ & $16$ & $20$ \\ \hline
Time (in seconds) & $0.63$    &  $1.74$   &   $4.85$   &  $12.27$    &   $137.63$   \\ \hline
\end{tabular}
\end{table}

\section{Numerical Results}
\label{Res}
In this section we conduct numerical experiments to assess the performance of our algorithm to offload computational tasks to the MEC server. 
Here, $g_{m}$ is modeled as $\left|h_{m}\right|^{2}/\zeta(\mathfrak d)$ where $h_{m}\sim\mathcal{CN}(0,1)$ denotes Rayleigh fading, and $[\zeta(\mathfrak d)]_{\mathrm{dB}} \triangleq 128.1 + 37.6 \log_{10}(\mathfrak d)$ is the distance-dependent path loss with $\mathfrak d$ denoting the distance between the transmitter and receiver in km~\cite{3GPP}.
We consider $B = 20$~MHz, which results in the noise power of $-92$~dBm. Note also that the channel gains are normalized w.r.t. the noise power at the MEC server. The distance of the users from the BS are given by $\{200, 250, 300,\ldots,200+50(M-1)\}$~m. Similarly, the time delay deadlines for the users are given by $\{D_{1},D_{1}+\Delta,D_{1}+2\Delta,\ldots,D_{1}+\Delta(M-1)\}$, where $D_{1}=0.5$~s and $\Delta=0.05$~s, unless stated otherwise. Note that, for simplicity, we have considered a constant difference between the users' deadlines, however, the proposed SCA-based algorithm holds good also for the case when the difference between the deadlines are random. Also, throughout this section, we assume that all of the users have the same transmit power budget, $P_{\mathrm t}$, i.e., $P_m = P_{\mathrm t}, \forall m \in \mathcal M$. We execute the proposed algorithm using CVXPY with MOSEK as the internal solver on a Linux PC with 7.5 GiB memory and Intel Core i5-7200U CPU. Note that in this section, we use the logarithm with base $2$ and thus the offloaded tasks are expressed in bits. For all of the figures in this section, the average values of offloaded bits are computed over the same set of $100$ independent channel realizations. Moreover, in Figs.~\ref{fig:Convergence},~\ref{fig:l_vs_Eth} and~\ref{fig:Deadline}, we consider $M=4$.

In Fig.~\ref{fig:Convergence}, where the numbers in the parentheses denote ($E_{\mathrm{th}}$ in mJ, $P_{\mathrm t}$ in dBm), we show the convergence behavior of the proposed algorithm for the NOMA-MEC system.  We also show the convergence of the corresponding OMA-MEC system for comparison. It can be noted from the figure that both NOMA and OMA systems converge within a small number of iterations. Moreover, for the given system parameters, the NOMA-based system results in a much larger number of offloaded bits within the same number of iterations, compared to the OMA-based counterpart. Also, in Table~\ref{TimeTable}, we show the average problem solving time for the NOMA-MEC system for $E_{\mathrm{th}}=5$~mJ, $P_\mathrm{t} = 5$~dBm and different number of users with number of iterations fixed at 20.

In Fig.~\ref{fig:l_vs_Eth} we report the average minimum number of offloaded bits as a function of energy threshold, $E_{\mathrm{th}}$ for given $P_{\mathrm t}$. From the figure, the superiority of the NOMA-based system is clearly evident over its OMA-based counterpart. More interestingly, for the case of NOMA-based system, the objective first increases for the small values of $E_{\mathrm{th}}$ (which we refer to as the \textit{energy-constrained} regime), and then saturates for large values of $E_{\mathrm{th}}$ (which we refer to as the \textit{power-constrained} regime). It is noteworthy that for two different NOMA systems with the same $E_{\mathrm{th}}$ but different $P_{\mathrm t}$, the performance in the energy-constrained regime remains the same. It can also be noted from the figure that compared to NOMA, the effect of easing the energy constraint is less pronounced in the OMA-MEC system.

To investigate the impact of number of users, in Fig.~\ref{fig:l_vs_M}, we plot the variation in the average minimum number of offloaded bits against the number of users. Note that in the figure legend, the numbers in the parentheses denote ($E_{\mathrm{th}}$ in mJ, $P_{\mathrm t}$ in dBm). It can be inferred from the figure that as the number of users, $M$, increases, the performance of both NOMA-based and OMA-based systems deteriorate. It is also evident that with an increase in the number of users, the performance gap between the NOMA-MEC and OMA-MEC systems also decreases. Note that for the NOMA-MEC system, as $M$ increases, the inter-user interference for the higher-indexed users also increases, reducing the achievable rate for such users.

To investigate the impact of users' delay requirements, in Fig.~\ref{fig:Deadline} we investigate the variation of number of offloaded bits with $\Delta$. Note that in the legend of the figure, the numbers in the parentheses denote ($E_{\mathrm{th}}$ in mJ, $P_{t}$ in dBm, $D_{1}$ in s). Remarkably, the performance of both NOMA-MEC and OMA-MEC systems increases linearly with a linear increase in $\Delta$.
The poor performance of OMA 
is due to the fact that $\mathrm{U}_{m}$ has to offload all its data only during the $m^{\mathrm{th}}$ time slot. Under the same circumstances NOMA is able to offload more data due to availability of additional time slot. Thus, NOMA-based MEC systems can be considered as a suitable candidate to achieve gains associated with edge servers under stringent delay constraints of users.
\section{Conclusion}
\label{Conc} In this correspondence, we have considered using NOMA to offload computational tasks for edge computing in a multi-user scenario. For the system model under consideration, we have formulated a non-convex optimization problem to maximize the minimum number of offloaded bits among the users, constrained by delay, transmit power and energy requirements. The original non-convex problem is solved using successive convex optimization strategy that leads to a provably convergent algorithm, each iteration of which can be solved efficiently by the SOCP solvers. The numerical results have showed that in NOMA systems with given power budget, the minimum number of offloaded bits first increases with energy budget in the energy-constrained regime, and then saturates in the power-constrained regime. It has also been confirmed that as the number of users increases, the performance gap between NOMA and OMA systems decreases. Our results have shown that \emph{for strict delay constraints}, NOMA substantially outperforms OMA-based approach in terms of maximizing tasks to be offloaded to the MEC server. 
\appendices
\section{Proof of Theorem~\ref{thm:1stApp}}
\label{Proof}

We can write $\ln(1+\frac{u}{v})=\ln(u+v)-\ln(v)$. Then, due to the concavity of $\ln(v)$ we have $\ln(v)\leq\ln(v_{0})+\frac{1}{v_{0}}(v-v_{0})=-1+\ln(v_{0})+\frac{v}{v_{0}}$. Next, we recall the following well-known inequality: $\ln(u)\geq1-\frac{1}{u}$, for $u>0$, where the equality occurs if and only if $u=1$. Therefore, it follows that $\ln(\frac{u+v}{u_{0}+v_{0}})\geq1-\frac{u_{0}+v_{0}}{u+v} \Rightarrow \ln(u+v)\geq1+\ln(u_{0}+v_{0})-\frac{u_{0}+v_{0}}{u+v}$. Using the preceding inequalities for $\ln(v)$ and $\ln(\frac{u+v}{u_{0}+v_{0}})$ yields $\ln(1+\frac{u}{v})\geq2+\ln(u_{0}+v_{0})-\ln(v_{0})-\frac{u_{0}+v_{0}}{u+v}-\frac{v}{v_{0}}$, and thus $x\ln(\!1+\frac{u}{v}\!)\!\geq\!x(\!2+\ln(\!1\!+\!\frac{u_{0}}{v_{0}}\!))-\frac{(u_{0}\!+\!v_{0})x}{u+v}-\frac{xv}{v_{0}}= x(2+\ln(1+\frac{u_{0}}{v_{0}}))-(u_{0}+v_{0})\frac{(x+1)^{2}-(x-1)^{2}}{4(u+v)} -\frac{(x+v)^{2}-(x-v)^{2}}{4v_{0}}$.

Since the term $\frac{(x-1)^{2}}{4(u+v)}$ is jointly convex in all variables involved, approximating the term $\frac{(x-1)^{2}}{4(u+v)}$ around $x_{0}$, $u_{0}$, and $v_{0}$ yields $\frac{(x-1)^{2}}{4(u+v)}\geq\frac{(x_{0}-1)^{2}}{4(u_{0}+v_{0})}+\frac{(x_{0}-1)}{2(u_{0}+v_{0})}(x-x_{0}) -\frac{(x_{0}-1)^{2}}{4(u_{0}+v_{0})^{2}}(u+v-u_{0}-v_{0}) = \frac{1-x_{0}}{2(u_{0}+v_{0})}+\frac{x(x_{0}-1)}{2(u_{0}+v_{0})}-\frac{(x_{0}-1)^{2}(u+v)}{4(u_{0}+v_{0})^{2}}$. Similarly we have $(x-v)^{2} \geq-(x_{0}-v_{0})^{2}+2(x_{0}-v_{0})(x-v)$.

Combining the preceding expressions for $x\ln(\!1+\frac{u}{v}\!)$, $\frac{(x-1)^{2}}{4(u+v)}$ and $(x-v)^{2}$ produces~\eqref{eq:LB}, which completes the proof.

\bibliographystyle{IEEEtran}
\bibliography{IEEEabrv,Data_Offloading_NOMA_MEC}
\end{document}